\documentclass[aps,prb,twocolumn,superscriptaddress]{revtex4-1}
\usepackage[utf8]{inputenc}   
\usepackage{graphicx}
\DeclareGraphicsExtensions{.png,.jpg,.eps}

\usepackage{xcolor}
\usepackage{amsmath}
\usepackage{amssymb,amsthm}

\usepackage{hyperref}
\hypersetup{colorlinks=true,linkcolor=blue,citecolor=blue,urlcolor=blue}

\newcommand{\ket}[1]{|#1\rangle}
\newcommand{\bra}[1]{\langle#1|}
\newcommand{\braket}[2]{\langle#1|#2\rangle}

\renewcommand{\Im}[1]{\operatorname{\mathbb{I}m}\left[#1\right]}

\begin{document}

\title{Anomalous magnetism in hydrogenated graphene}
\author{N. A. Garc\'ia-Mart\'inez}
\affiliation{QuantaLab, International Iberian Nanotechnology Laboratory (INL), Braga, Portugal}

\author{J. L. Lado}
\affiliation{QuantaLab, International Iberian Nanotechnology Laboratory (INL), Braga, Portugal}

\author{D. Jacob}
\affiliation{Max-Planck-Institut f\"ur Mikrostrukturphysik, Weinberg 2, 06120 Halle, Germany}

\author{J. Fern\'andez-Rossier}
\affiliation{QuantaLab, International Iberian Nanotechnology Laboratory (INL), Braga, Portugal}
\affiliation{Departamento de Fisica Aplicada, Universidad de Alicante, San Vicente del Raspeig, 03690 Spain}

\date{\today}

\begin{abstract}
We revisit the problem of local moment formation in graphene due to chemisorption of individual atomic hydrogen or other analogous sp$^3$ covalent functionalizations.
We describe graphene with the single orbital Hubbard model, so that the H chemisorption is equivalent to a vacancy in the honeycomb lattice. In order to circumvent artifacts related to periodic unit cells, we use either huge simulation cells of up to $8\times10^5$ sites, or an  embedding scheme that allows the modeling of a single vacancy in an otherwise pristine infinite honeycomb lattice. We find three results that stress the anomalous nature of the magnetic moment ($m$) in this system. First,  in the non-interacting ($U=0$), zero temperature ($T=0$) case, the $m(B)$ is a continuous smooth curve with divergent susceptibility, different from  the step-wise constant function found for a single unpaired spins in a gapped system. Second, for $U=0$ and $T>0$, the linear susceptibility  follows a power law $\propto{T}^{-\alpha}$ with an exponent of $\alpha=0.77$ different from conventional Curie's law. For $U>0$, in the mean field approximation, the integrated moment is smaller than $m=1\mu_B$, in contrast with results using periodic unit cells.
These three results highlight that the magnetic response of  the local moment  induced by sp$^3$ functionalizations in graphene is different both from  that of local moments in gaped systems, for which the magnetic moment is quantized and  follows a Curie law,  and from Pauli paramagnetism in  conductors, for which a linear susceptibility can be defined at $T=0$.
\end{abstract}

\maketitle

\section{Introduction}
Whether or not the addition of an impurity atom into an otherwise non-magnetic crystal results in the formation of local moments is one of the central problems in condensed matter physics.\cite{anderson1961localized}
In the cases when the host is either a conductor with a well defined Fermi surface, or an insulator with the Fermi energy inside an energy gap, the problem is well understood.
In the former case, the formation of a local moment is controlled by the competition between the addition energies of the impurity levels and their broadening, due to quasi-particle tunneling in and out of the impurity\cite{anderson1961localized,Daybell1968}. This quantity depends on the density of states of the energy of the localized level.
In contrast, when the impurity level lies inside a gap, as it happens for donors in semiconductors, the unpaired electronic spin behaves like a paramagnetic center with $S=1/2$.\cite{Slichter1955}

The hydrogenation of graphene has attracted interest for various reasons. In
the dense limit, it leads to the opening of a large
band-gap.\cite{elias2009,Sofo2007} In the dilute limit, chemisorbed hydrogen
was predicted to create a $S=1/2$ local moment
\cite{Duplock2004,yazyev2007,Boukhvalov2008,Palacios2008,Yazyev2010,santos2012} associated
with the formation of an $E=0$, or mid-gap, state that hosts an unpaired electron.
This led early on to propose hydrogenated graphene as a magnetic material with
spin-dependent transport properties\cite{Soriano2010,Leconte2011,Soriano2011}
apt for spintronics in graphene.\cite{han2014graphene} Recent scanning
tunneling microscope experiments\cite{gonzalez2016atomic} match the computed
theoretical density of states as a function of both energy and position, that
shows a split resonance close to the Dirac point, providing thereby indirect
evidence for local moment formation. 
It also has been proposed that the small local lattice distortion induced by
the $sp^3$ hybridization enhances the effect of spin-orbit
interaction.\cite{Castro2009,gmitra13} This has been proposed as the
explanation for the observation of large spin Hall angles in hydrogenated
graphene.\cite{balakrishnan2013colossal} Resonant scattering with the zero mode
resonance has also been considered as a source of enhanced quasi-particle spin
relaxation in graphene.\cite{Wojtaszek2013,Kochan2014,soriano2015}

Chemisorption of atomic Hydrogen in graphene structures entails the formation
of a strong covalent bond between the carbon $\pi$ orbital and the hydrogen $s$
orbital. The pair of bonding-antibonding states lies far from the Fermi energy.
This picture is valid not only for chemisorption of atomic hydrogen, but for a
large variety of other  $sp^3$ adsorbates. \cite{santos2012}
Effectively, the result of this  hybridization is to remove both one electron
and one state from the $\pi$ cloud, which justifies modeling $sp^3$
functionalization with the one orbital tight-binding model with the removal of
one atomic site, and 1 electron, in the honeycomb lattice\footnote{This vacancy
in the effective model is different from a real vacancy in graphene, which
entails the formation dangling bonds from the $sp^2$ orbitals, in addition to
the removal of the $\pi$ orbital as well}.
The removal of a site in the honeycomb lattice breaks sub-lattice symmetry and produces a zero energy state.\cite{Pereira2006,kumazaki2007,Wehling2007,Pereira2008,Palacios2008}

In gapped graphene structures, such as graphene with a spin-orbit gap,
\cite{Gonzalez2012} graphene nanoribbons,\cite{Palacios2008} or a planar
aromatic hydrocarbon molecule, the formation of an in-gap $E=0$ state due to
$sp^3$ functionalization trivially leads to a local moment formation with
$S=1/2$, very much like it happens for acceptors and donors in semiconductors.
The zero energy state is singly occupied by one electron
that occupies a bound state.
The rest of this manuscript is devoted to study the case of a $sp^3$ chemisorption in infinite gapless graphene. In that situation, the formation of a local moment is not warranted. Whereas the $E=0$  state appears exactly at the energy where the DOS vanishes, a finite DOS due to the Dirac bands is infinitesimally close, and its wave function, described with $\psi\propto\frac{1}{x\pm i y}$ is a quasi-localized non-normalizable state.\cite{Pereira2006}
The DOS of the resonant state diverges at $E=0$ and hence lacks a Lorentzian lineshape.   
This situation is genuinely different from conventional magnetic impurities in metals, that are coupled to a bath with finite DOS and show conventional Lorentzian lineshapes and it is also different for the situation in which the resonant state lies inside a proper gap.

Previous works usually address this problem by using supercells, where translation symmetry is preserved, hence considering not a single defect but a periodic array of them.\cite{Duplock2004,Boukhvalov2008,Yazyev2010,Sofo2012,gonzalez2016atomic}
In this case the bands of such a system will always present a gap which at half filling, results in a quantized $m=1\mu_B$ magnetic moment since all the valence bands would be doubly occupied but only one of the two states in the gap would be occupied.
While this approach has been proven very useful in many studies, it does not strictly solve the problem of a single impurity in an otherwise pristine graphene sheet.
To tackle this problem we use two different methodologies. On the one hand, we
make a Green's function description of the defected region embedded into an
infinite pristine crystal which yields an exact description, but results in
computationally relatively expensive calculations. On the other hand, we use
the kernel polynomial method which allows the calculation of spectral
properties for huge systems (800000 atoms) in a computationally efficient way,
but has the drawback of lower resolution in energy.

The rest of the paper is organized as follows. Section~\ref{sec:GF} is devoted
to present the methods, i.e. the formalism to treat a single impurity in an
infinite system. In Sec.~\ref{sec:Phys} we study the physical properties of a
single $sp^3$ chemisorption on graphene including the effect of an external
magnetic field with and without temperature. Sec.~\ref{sec:MF} is devoted to
the study of electron-electron interactions, and finally in
Sec.~\ref{sec:Concl} we summarize our work.

\section{Methods}
\label{sec:GF}
In the following, we describe graphene with a one orbital tight-binding model.  Electron-electron interactions are described within the Hubbard approximation.  Thus, the energy scales in the Hamiltonian are the first neighbor hopping $t$ and the Hubbard on-site repulsion $U$. When an external magnetic field is introduced it is considered to be an in-plane magnetic field $B$ coupled only to the spin degree of freedom.
The effect of $sp^3$ hybridization is included by the removal of both of a site in the honeycomb lattice and one electron, as discussed above.

We use two different techniques to tackle the problem of a single impurity in
pristine graphene. The first one consists in the calculation of the (exact)
Green's function for a region close to the defect by means of the Dyson
equation, using an embedding method described below.
The second one is the kernel polynomial method, that allows the calculation of spectral properties of extremely large systems with minimal computational effort.

\subsection{The embedding technique}
We first present a general method to study single impurities in infinite systems, from now on referred as embedding technique, devised earlier by one of us in a similar context.\cite{Jacob2010orbital}
Since a single defect in an infinite system breaks the translation symmetry, we cannot use a Bloch description. Instead, we describe the system in terms of Green's functions, making use of the Dyson equation.
\begin{figure}[t!]
\centering\includegraphics[width=0.5\textwidth]{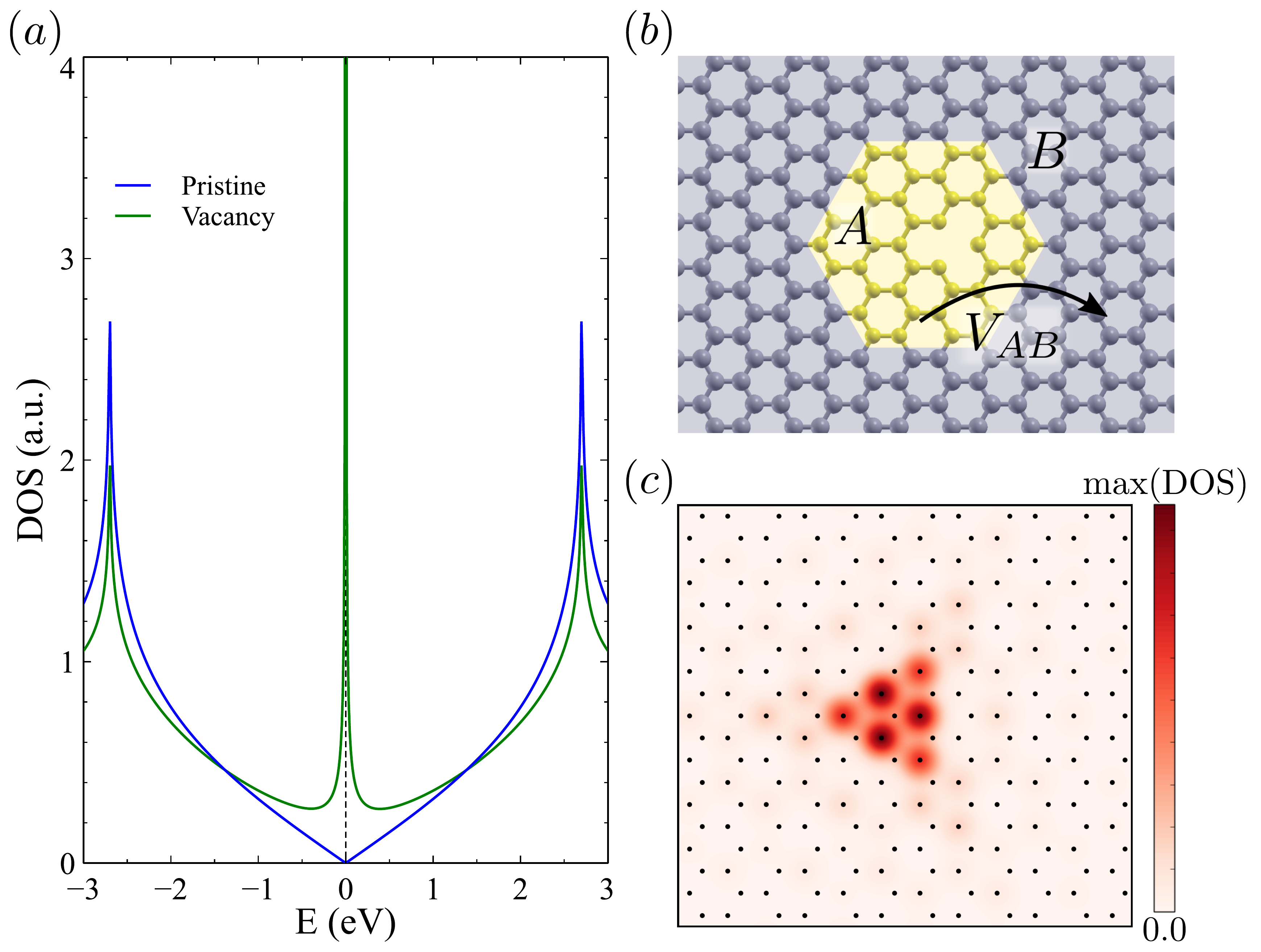}
\caption{$(a)$ Total density of states of a single vacancy in an infinite graphene sheet. A divergence in the density of states appears at $E=0$ when the vacancy is introduced. $(b)$ Scheme of the division of the system into a defected unit cell and a pristine environment. $(c)$ Local Density of States for the zero energy state related to a vacancy in graphene. Side by side we can compare the calculations for two unit cells with different geometry. The vacancy is depicted as a white circle. As expected the spatial distribution of this state is located in the 3-6 closest atoms to the vacancy.}
\label{regions}
\label{DOS}
\vspace{-5pt}
\end{figure}
We start by dividing the system into two regions, a central unit cell $A$ containing the defect, and the rest of the system $B$, containing everything else, as depicted in Fig.~\ref{regions} (b).
The Hamiltonian of the whole (infinite) system can then be written in terms of the two separated contributions, one arising from each isolated region, $H_0$, and the other arising from the coupling between the two regions $W$:

\begin{equation}
  H = H_{0} + W = \left(\begin{array}{cc}
   H_{A} &  0  \\
    0     & H_{B}
  \end{array}\right)+
  \left(\begin{array}{cc}
    0 & V_{AB} \\
   V_{BA} & 0
  \end{array}\right)
\end{equation}

The Green's function corresponding to region $A$ can be written (exactly) as:

\begin{equation}
  G_{A}(E) 
  = \frac{1}{E +i\eta - H_{A}-\Sigma_{AB}(E)}
  \label{Gdyson}
\end{equation}
where the \emph{embedding self-energy} $\Sigma_{AB}$ can be calculated from the Green's function of region $B$ 
$g_B(E)=(E+i\eta-H_B)^{-1}$, as $\Sigma_{AB}(E)=V_{AB}\;g_B(E)\;V_{BA}$.
For numerical reasons $\eta$ has to be finite but we checked that the results do not depend on its exact value. We found that $\eta=0.001$ offers a good combination of precision in energy while keeping the convergence time of the Dyson equation reasonable.
In general the Green's function $g_B(E)$ for region $B$
are not straightforward to calculate, as $g_B(E)$ describes the Green's function of an infinite system without translation symmetry, on account of the missing region $A$.
However, the calculation of $\Sigma_{AB}$ is made possible when we consider two facts. First, $\Sigma_{AB}$ does not depend on whatever is in region $A$, second, equation \eqref{Gdyson} holds true for a pristine system with translation invariance, that permits to compute $G_A(E)$, and evaluate $\Sigma_{AB}=E- H_A-(G_A(E))^{-1}$.

The evaluation of $G_A(E)$ is now done by dividing the infinite pristine crystal system into periodic supercells $A'$ of the same size and shape as the defect region $A$.
The Green's function of region $A'$ in the perfect crystal can thus be calculated by integrating the $\vec{k}$-dependent Green's function in the whole Brillouin zone
\begin{equation}
  G_{A'}(E) = 
\frac{1}{(2\pi)^2}
\int_{\text{BZ}} (E+i\eta-H(\vec{k}))^{-1} d^2\vec{k}
  \label{Gbloch}
\end{equation}
with $\vec{k}$ {the Bloch wavevectors}
and $H(\vec{k})$ the Bloch Hamiltonian for the pristine host
crystal. The final expression for the self-energy $\Sigma_{AB}$ reads:
\begin{equation}
  \Sigma_{AB}(E) = E+i\eta - H_{A'} - 
\frac{1}{(2\pi)^2}
\left ({\int_{\text{BZ}} 
(E+i\eta-H(\vec{k})})^{-1}d^2\vec{k} \right )^{-1}
\label{selfenergy}
\end{equation}
where $H_{A'}$ describes a region with the same dimensions than the original defective region $A$, but without the defect(s). This is a general procedure and can be applied for multi-band Hamiltonians. As long as the dimensions of the pristine and the defected Hamiltonian are the same, it can deal with more than one defect without computational overhead. Notice that this method does not require the analytic evaluation of the host crystal Green's function necessary in a recently proposed method\cite{settnes2015}, and can be applied to a very large class of systems, including superconductors.\cite{lado2016}

The combination of equations \eqref{Gdyson} and \eqref{selfenergy} allows the computation of the Green's function of the defective area, $G_A$, embedded in an otherwise pristine crystal as shown in Fig.~\ref{regions}. The density of states (DOS) of an atom $i$ in region $A$ can then be calculated from the imaginary part of the Green's function as
\begin{equation}
  \rho_{i}(E) = -\frac{1}{\pi}\Im{G_{i,i}(E)}
  \label{eq:DOS}
\end{equation}
where $G_{i,i}$ is the diagonal matrix element $(i,i)$ of the Green's function. Summing over the contributions from all atoms $i$ in region $A$, the total DOS of region $A$ is obtained.

In Fig.~\ref{DOS} we show the results of the method for the case of a single vacancy in the honeycomb lattice. Fig.~\ref{DOS} (a) shows the density of states both for pristine graphene, that shows the characteristic $\rho \propto |E|$ around the Dirac point,\cite{Katsnelson2012} and for the defective case, that presents a diverging zero energy resonance. The embedding method permits also the calculation of the local density of states as shown in Fig.~\ref{DOS} (c), where we show the map of the density of states evaluated at $E=0$, finding that the main contribution for this state comes from the 3-6 nearest neighbors to the vacancy that belong to the sublattice opposite to the one of the missing site.\cite{Pereira2006,kumazaki2007,Wehling2007,Pereira2008,Palacios2008}
Of course, for the case of a non-interacting single vacancy, this problem can be dealt with using the standard T matrix theory.\cite{libro-Economou,Gonzalez2012,Wehling2007}. The embedding method shows its added value when it comes to treat several vacancies or when interactions are included, as we discuss now.

\subsection{Mean field Hubbard model }
The Hubbard term acting on every site $i$ reads:
\begin{equation}
  H_{U} = U \sum_i n_{i,\uparrow}\,n_{i,\downarrow}
\end{equation}
where $n_{i\sigma}$ is the standard number operator for site $i$ with spin $\sigma$.
Exact solutions of this model are, in general, not possible so that we use a mean field approximation:
\begin{equation}
  H_{U} \approx \sum_i
  U \left[\langle n_{i,\uparrow} \rangle n_{i,\downarrow} +
    \langle n_{i,\downarrow} \rangle n_{i,\uparrow}  -
    \langle n_{i,\downarrow} \rangle \langle n_{i,\uparrow} \rangle
  \right]
\end{equation}
where $\langle n_{i,\sigma} \rangle$ stand for the expectation values of the number operators computed with the eigenstates of the mean field Hamiltonian.
Of course, this is a non-linear problem that is solved self-consistently.  Here, this is done in combination with the two dimensional embedding technique, which is formally similar to the one dimensional case.\cite{munoz2009}
In this approach, the occupations in the external region $B$ are frozen to $\langle n_{i,\uparrow}\rangle= \langle n_{i,\downarrow}\rangle=\frac{1}{2}$.  In contrast, the expected values of the defective cell are calculated by self-consistent iteration.

In a first step, we assume a random guess spin polarization and then compute the expected values of the spin operators $\langle n_{i,\sigma} \rangle$ by integrating the DOS up to the Fermi energy
\begin{equation}
 \langle n_{i,\sigma}\rangle= \int_{-\infty}^{E_F} \rho_{\sigma}(E) dE
 \label{occ}
\end{equation}
which defines a new Hamiltonian for region $A$,$H_A \rightarrow \bar H_A + H^{MF}_U$
, including the mean field Hubbard term.\cite{Fernandez2007,Palacios2008} Notice that the numerical integration of eq. \eqref{occ} is much more efficiently done in the complex plane using Cauchy's integral theorem. Also it is important to notice that even when the Hamiltonian for the region $A$ will change over the self-consistent iterations the self-energies will not since they do not depend on what is inside of said region.
This procedure is iterated until a self-consistent solution is found.

The magnetic moment is calculated as the difference of the expected values of each spin densities.
\begin{equation}
\langle m(i)\rangle \equiv g\mu_B\frac{\langle n_{i,\uparrow}\rangle-\langle n_{i,\downarrow}\rangle}{2}
\label{sz}
\end{equation}

There is a trade off between computational cost, due mainly to the size of the region $A$, and the accuracy of the description of the semi-localized nature of the induced magnetism. The role of the chosen size for region $A$ is discussed bellow.

\subsection{The kernel polynomial method}
The Kernel polynomial method\cite{Weisse2006} is a spectral method that allows to calculate spectral properties of very large matrices without explicit diagonalization or inversion of the matrix.
This makes the method especially suitable for very large systems described by sparse Hamiltonians, as is the case for the first neighbor hopping model for the honeycomb lattice, considered here.
In our case, we set up the Hamiltonian for an extremely large graphene island, with a single vacancy in the center.

{The Chebyshev polynomials form a complete basis in the function space, so that they can be used as a basis to expand any well behaved function} $f(x)$ for $x\in (-1, 1)$
The method consists in expanding the density of states in $N$ Chebyshev polynomials $T_n(x)$,  that are calculated using $T_0(x)=1$, $T_1(x)=x$ and the recursive relation
\begin{equation}
T_{n+1} (x) = 2 x T_n(x) - T_{n-1} (x)
\end{equation}
valid for $x\in (-1,1)$.

The first step in the method is to  scale the Hamiltonian $H \rightarrow \bar H=\sum_k\bar E_k \ket{k}\bra{k}$
so that all the eigenstates $\bar E_k$ fall in the interval $\bar E_k \in (-1,1)$.  The density of states as a function of the scaled energy, at site $i$,  is expressed as
\begin{equation}
\rho_i(\bar E) = \frac{1}{\pi \sqrt{1-\bar E^2}}
\left (\bar \mu_0 + 2 \sum^{N-1}_{n=1} \bar \mu_n T_n (\bar E)
\right )
\label{KPM}
\end{equation}
The coefficients $\bar \mu_n$ are
the modified coefficients of the expansion,
\begin{equation}
\bar \mu_n = g^N_n \mu_n
\end{equation}
that are obtained using the Jackson kernel\cite{jackson1912approximation}
\begin{equation}
g_n^N =
\frac{(N-n-1)\cos \frac{\pi n}{N+1} + \sin \frac{\pi n}{N+1}
\cot \frac{\pi }{N+1}}{N+1}
\end{equation}
that improves the convergence of the expansion. The original
Chebyshev coefficients are calculated as a conventional
functional expansion
\begin{equation}
\mu_n = \int_{-1}^{1} T_n(\bar E) \sum_k \delta (\bar E-\bar E_k) |\langle k | i \rangle |^2
= \langle i | T_n(\bar H) | i \rangle
\label{mun}
\end{equation}
with $| i \rangle$ the wave function localized in site $i$. Importantly, the
second equality in eq. (\ref{mun}) relates $\mu_n$ to an expression where the
eigenstates $|k\rangle$ of $H$ are absent. Thus, diagonalization of $H$ is not
necessary and  the computation of the the $\mu_n$ coefficients only requires
calculating an overlap matrix element involving $T_n(H)$.
The Chebyshev recursion relation allow to write down the $\mu_n$ coefficients in term of the overlaps with the vectors $\ket{\alpha_n}$
\begin{equation}
\mu_n =
\braket{\alpha_0}{\alpha_n}
\end{equation}
generated by the recursion relation
\begin{equation}
\begin{aligned}
\ket{\alpha_0} = \ket{i}  \\  
\ket{\alpha_1} = \bar{H} \ket{\alpha_0} \\
\ket{\alpha_{n+1}} = 2\bar{H} \ket{\alpha_n}- \ket{\alpha_{n-1}}
\end{aligned}
\end{equation}
In our case, we will choose a state, $\ket{i}$, localized in the first neighbor
of the carbon with the hydrogen ad-atom.
To calculate the previous coefficients we only need matrix vector products,
so that the scaling is linear with the size $L$ of the system, in contrast with
the $L^3$ scaling for exact diagonalization.
Our calculations are performed in a graphene island with 800000 atoms,
taking a expansion with $N=10000$ polynomials.

\section{Non interacting zero temperature magnetization}
\label{sec:Phys}
In this section we study the spin polarization in the neighborhood of a $sp^3$ defect, driven by an external in-plane magnetic field coupled to the electronic spin, at zero temperature and in the non-interacting limit $U=0$.
In a gapped graphene system, this problem is straightforward. At $T=0$,  the
spin density would be dominated by the contribution of the only singly occupied
state, the $E=0$ mid-gap state, whose wave function we denoted by $\psi_0(i)\equiv \braket{i}{\psi_0}$.
The zero temperature magnetization in an atom $i$ would be given by:
\begin{equation}
m_i(B) = g\mu_B \left(\Theta(B)-\frac{1}{2}\right) |\psi_0(i)|^2
\end{equation}
where $B$ is the magnitude of the magnetic field, $\Theta(B)$ is the step function, $g\simeq2$ is the gyromagnetic ratio and $\mu_B$ is the Bohr magneton. The local magnetization $m_i$ is stepwise constant, and discontinuous at $B=0$, $m_i(0^+)-m_i(0^-)= g\mu_B |\psi_0(i)|^2$.
It is apparent that the total moment $M=\sum_i m_i$ integrates to $M=\pm \frac{g\mu_B}{2}$ on account of the normalization of the wave function of the mid-gap state. This result holds true as long as the Zeeman energy $\mu_B B$  is smaller than the gap of the structure.

\begin{figure}[t!]
  \centering
  \includegraphics[width=0.5\textwidth]{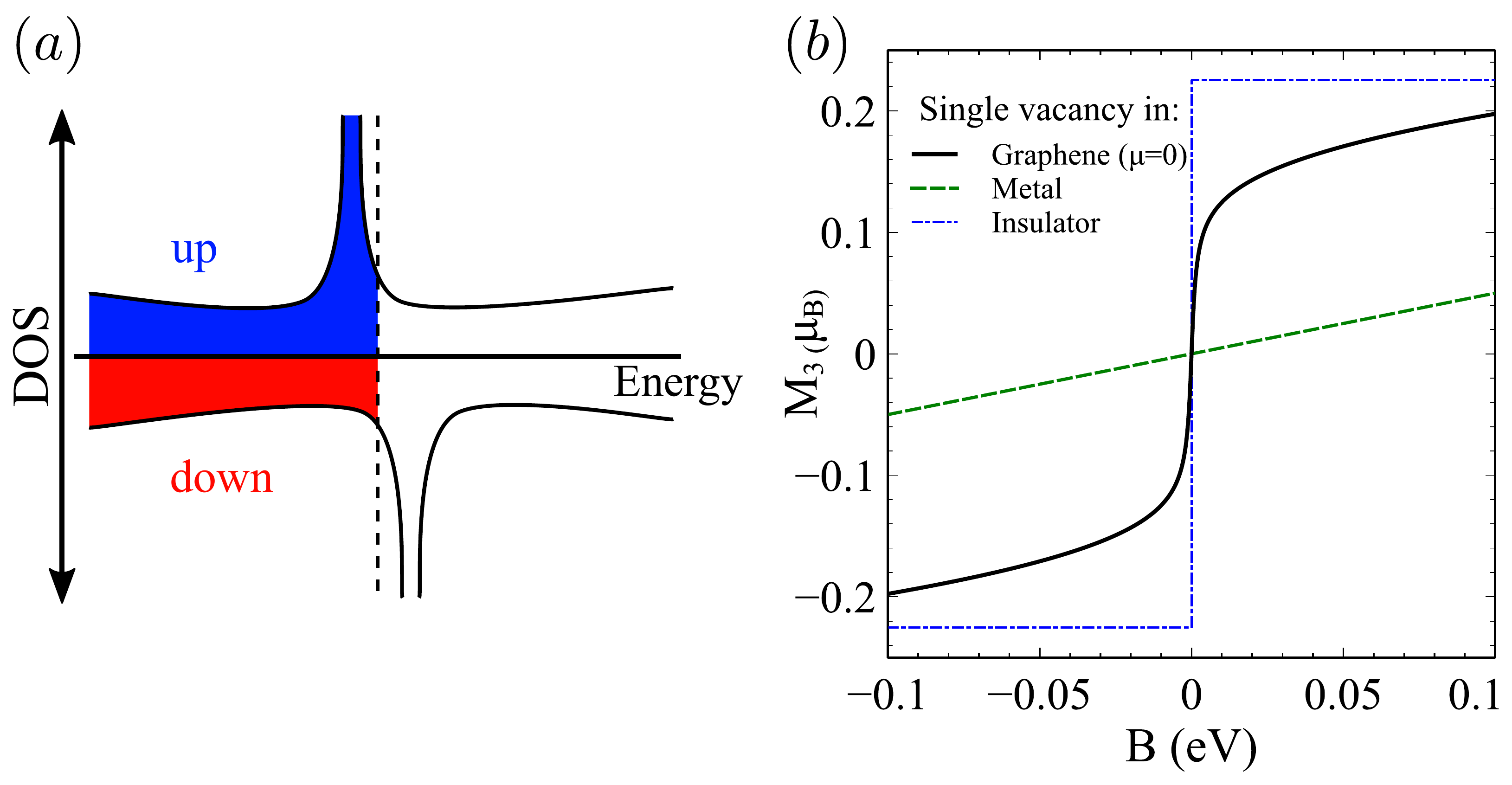}
  \caption{$(a)$ Sketch for the Zeeman split  DOS associated to  graphene with an individual $sp^3$ functionalization. Panel $(b)$ shows the magnetization of the first 3 neighbors of the defect as a function of the applied magnetic field for a hydrogen atom in a graphene quantum dot (blue) and a single hydrogen atom in pristine graphene (black), the green line is the result for a conventional metal, modeled by graphene with the chemical potential well above the Dirac point.}
  \label{mb}
\end{figure}
We now study what happens in the case of infinite pristine graphene, for which there is no gap, and we cannot define a normalized zero energy state. For that matter, we compute the density of states of the system using the Green's function embedding approach 
This methods is trivially adapted to include the Zeeman splitting that introduces a rigid spin dependent energy shift $\pm\mu_B B$. This symmetric shift allows the expression of all the spectral functions for each of the spin channels in terms of the spinless Green's function, $G^\sigma(E)=G(E-\sigma \mu_BB)$, with $\sigma= \pm 1$.


The results for the magnetization of the three first neighbors of a $sp^3$ defect in an otherwise pristine graphene are shown in Fig.~\ref{mb} for a single $sp^3$ defect in two scenarios. A gapped finite size graphene hexagonal island with armchair edges, resulting in the expected step-wise response, a single defect in otherwise pristine gapless graphene.
In both cases we plot  the magnetization of the three atoms closest to the vacancy,
$M_3=\sum^3_{i=1}m_i$, that gives the dominant contribution to the
defect-induced local moment. {The result for the paramagnetic response of a
metal is included in Fig.~\ref{mb} for comparison with a standard case}.

The most prominent feature of the obtained results is the fact that the
$M_3(B)$ curve is not stepwise constant for the defect in infinite graphene, in marked contrast with the case of the defect in a gapped island.
This difference shows the qualitatively different behavior of the zero mode in gapless infinite graphene, compared to the standard case of an in-gap truly localized state. The \emph{continuous variation} of the magnetic moment can be related to the fact that the zero mode has an intrinsic line-width that reflects the lack of a gap to host a true localized state.


\section{Finite temperature susceptibility}
\label{sec:Temp}
We now discuss the effect of temperature on the non-interacting $m(B)$ curve. The only effect of temperature is to smear out the occupation of the one-particle levels, so that the expected value of the local magnetization has to include now excited states.

\begin{figure}[t!]
  \centering
  \includegraphics[width=0.5\textwidth]{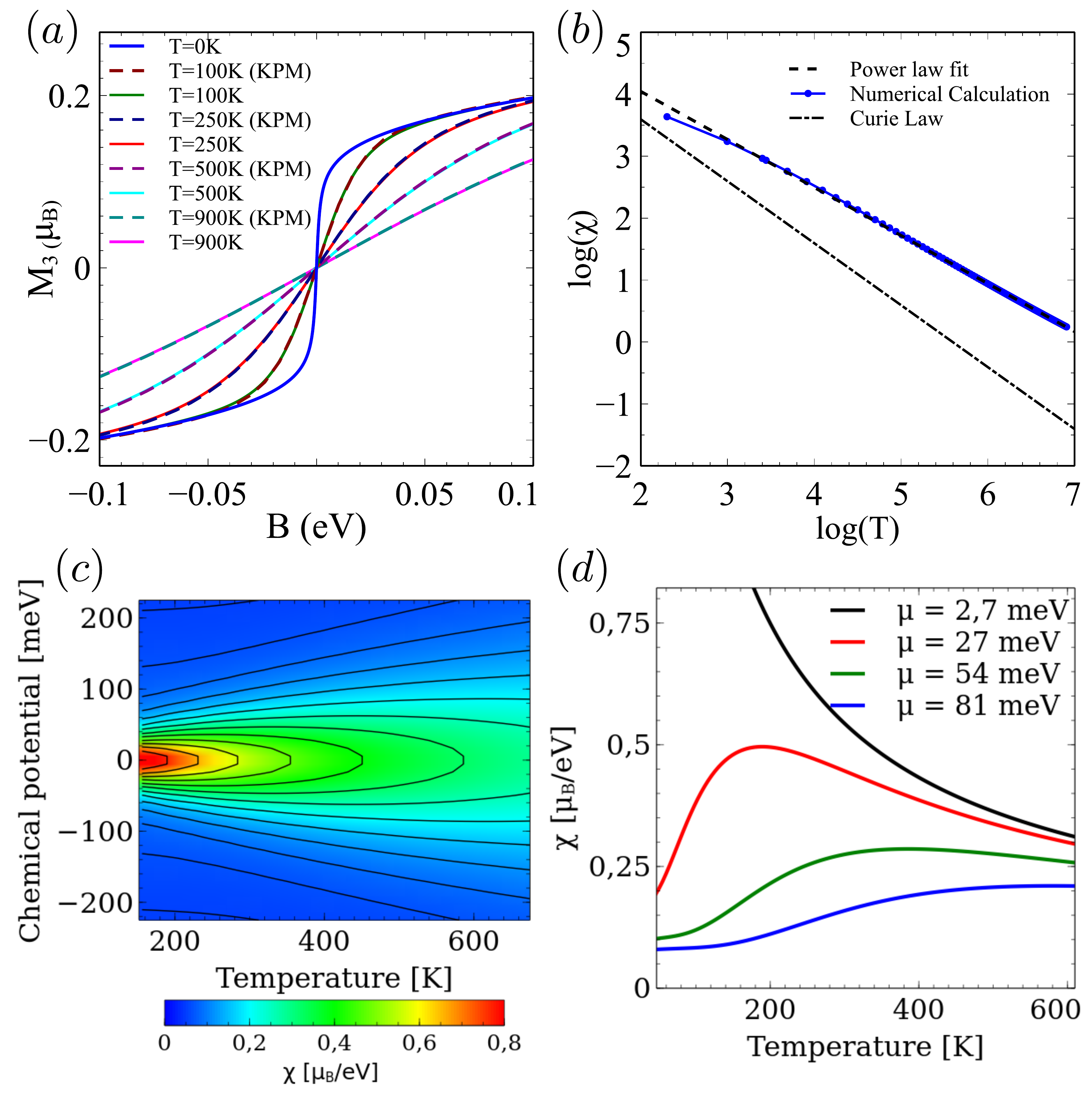}
  \caption{$(a)$ Magnetization of the first 3 neighbors of the vacancy as a
    function of the applied magnetic field for different temperatures, dashed lines are calculated using the KPM and continuous lines are calculated using the embedding method. $(b)$ Temperature dependence of the susceptibility in comparison with Curie's Law. $(c$-$d)$ Dependence of the susceptibility with the doping and the temperature, showing that for some values of $\mu$ there is a non-monotonous behavior of the susceptibility with temperature.}
  \label{mag_temp}
\end{figure}

To calculate the magnetization in a site $i$  as a function of the magnetic field and the temperature we just need to compute the difference in the occupation of the spin-up and spin-down density of states weighted with the Fermi-Dirac distribution function. Thus, the local magnetic moment is given by $m_i = g\mu_B \langle s_z(i) \rangle$ with
\begin{equation}
       \langle s_z(i) \rangle = \frac{1}{2}
      \int^{\infty}_{-\infty}\left[
      \rho_{i\uparrow}(E)-\rho_{i \downarrow}(E)
      \right] f(E,T) dE
\label{mag_b_t}
\end{equation}
where $f(E,T)$ is the Fermi Dirac distribution and $\rho_{i\sigma}(E)$ is the
spin resolved density of states. The resulting magnetization of the three
closest atoms, $M_3$ is shown in Fig.~\ref{mag_temp} (a), computed with two
different approaches, the embedding method for a single $sp^3$ defect in
infinite graphene and the kernel polynomial method for a single defect on a
finite size island with a very large number of atoms.
It is apparent that both methods give identical results in the chosen range of temperatures.

In order to highlight the anomalous behavior of the magnetic  moment associated to an individual $sp^3$ defect, we focus on the spin susceptibility, defined as
\begin{equation}
  \chi(T) = \left.\frac{\partial m (T)}{\partial B}\right|_{B=0}
\label{susceptibility}
\end{equation}
For $T=0$ the results of the previous section show that this quantity diverges, both for the gapped and gapless cases. Here we study the dependence of $\chi(T)$ as a function of temperature $T$. For a conventional local moment,  the zero field susceptibility follows the Curie law $\chi(T) \propto T^{-1}$. This result holds true, based on very general considerations, for any spin governed by the Hamiltonian $g\mu_B  \vec{S}\cdot\vec{B}$ as well as any classical magnetic moment $\vec{M}$ governed by the interaction energy $-\vec{M}\cdot\vec{B}$. In particular, the Curie susceptibility of a single electron in an in-gap level will follow a Curie law.

Numerical derivation of the results of  Fig.~\ref{mag_temp} (a) allows the calculation of $\chi(T)$, shown in Fig.~\ref{mag_temp} (b) in a Log-Log representation.
It is apparent that the spin susceptibility for the $sp^3$ defect on graphene
\emph{does not follow} the Curie law.
In particular, we obtain a high temperature power law dependence $\chi \propto T^{-\alpha}$, with $\alpha\sim0.77$, in comparison with the conventional $\alpha=1$.
This exponent reflects, again, the anomalous nature of the $sp^3$ local moment in infinite graphene, in marked contrast with the behavior of the same chemical functionalization in a gapped graphene structure.
Interestingly, $\chi(T)$ has been measured\cite{Nair2012} for defective graphene obtaining a Curie law dependence, probably because the samples used are in fact nanoflakes with small confinement gaps that permit the existence of in-gap states with quantized spins.

Further insight into the magnetic properties of this system is obtained by considering the dependence of the magnetic susceptibility as a function of the graphene chemical potential, that could be modified by gating doping,\cite{nair2013dual} as shown in Fig.~\ref{mag_temp} (c,d).
The maximal local susceptibility is obtained at half-filling and small temperatures where it monotonically decreases both with temperature and doping.
In comparison, in the case of slightly doped samples, the magnetic
susceptibility can either increase or decrease as a function of the
temperature, showing a maximum at a doping-dependent temperature. The previous
behavior can be understood as a crossover from the impurity in an insulator to
the metallic regime.
Importantly, the local maximum implies that graphene doping introduces an energy
scale that determines the temperature for the impurity-metal crossover. In the
case of heavily doped samples ($\mu = 81$ meV),
the susceptibility grows monotonically with temperature,
signaling the conventional metallic regime.

\section{Effect of interactions}
\label{sec:MF}
We now study the effect of electron-electron interactions in the formation of local magnetic moments associated to the $sp^3$ functionalization. A single unpaired electron in an in-gap energy level has a spin $S=1/2$ (equivalent ot a magnetic moment $m=1\mu_B$). In the current study this is not the case, the $E=0$ resonance is embedded in a region with finite DOS (except in just one point, the Dirac energy) and the existence of an emergent magnetic moment follows the Stoner criterion $U\rho(E_F)=1$. For graphene with a single $sp^3$ defect the diverging nature of $\rho(E)$ at $E=0$, the presence of arbitrarily small interactions gives rise to a local moment.
However, because of the coupling of the mid-gap state to a continuum of states (the linear bands of graphene) it is not obvious a priori whether or not the local moment should have a quantized $S=1/2$ spin.
In the following we address this issue, using the mean field approximation for the Hubbard model and the embedding technique as discussed in previous sections. The Hubbard model, while is not certainly the complete description of such a system, offers a simple model to gain an insight on the possible magnetic solutions for the systems.

\subsection{Local magnetic moment}
\begin{figure}[t!]
  \centering
    \includegraphics[width=0.5\textwidth]{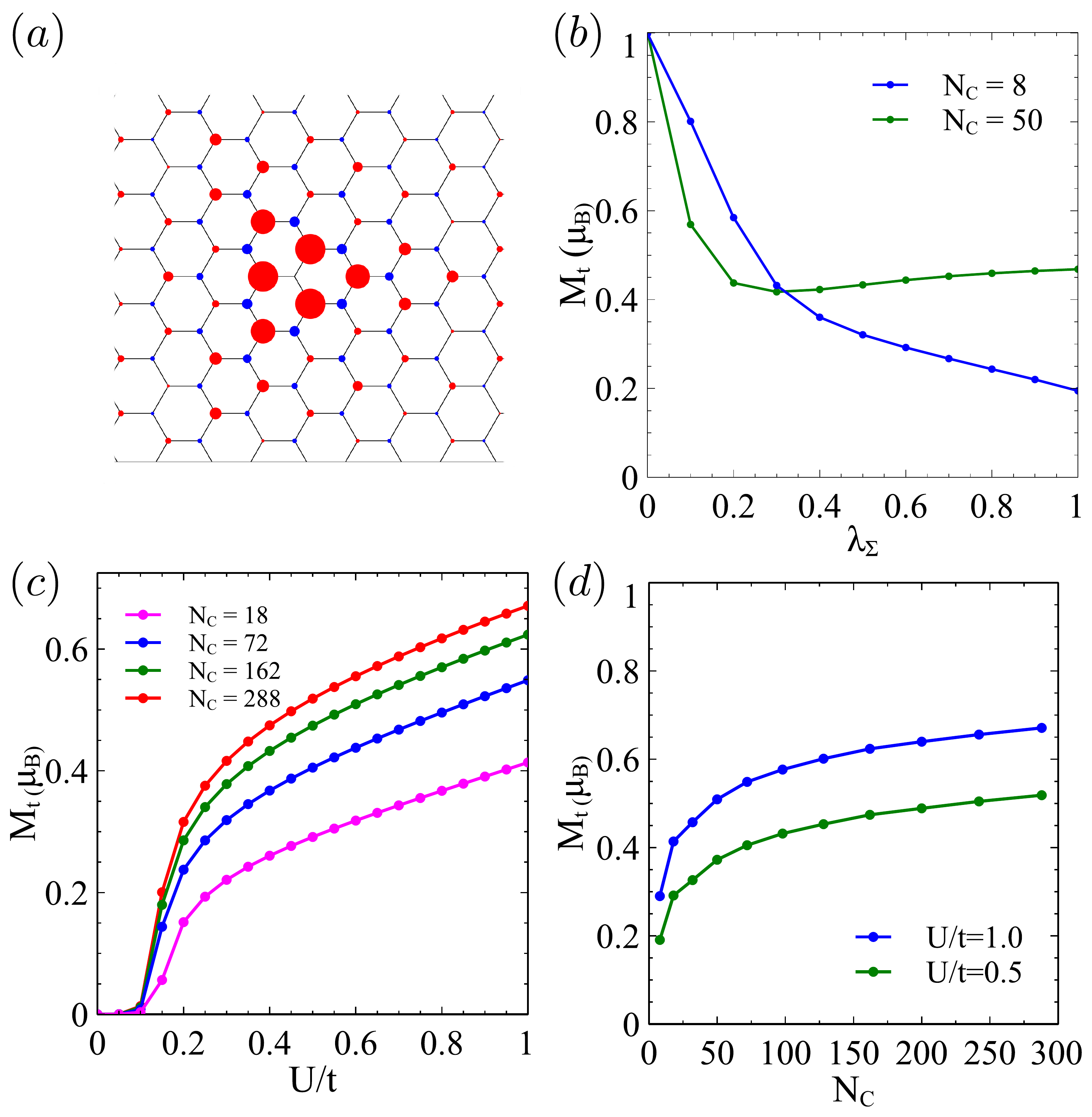}
  \vspace{-15pt}
  \caption{$(a)$ Magnetization of an individual $sp^3$ functionalized system, calculated within the mean field  Hubbard  approximation. $(b)$ Total magnetization of the defected region as a function of its coupling to the rest of the otherwise pristine system. $(c)$ Total magnetization of the defected region as a function of the Hubbard $U$ for different sizes of the unit cells. $(d)$ Total magnetization as a function of the size of the unit cell for two $U$ values, notice that the magnetization is far from the expected $m=1\mu_B$ value.}
  \label{MF}
\end{figure}

In general, the results of the mean field calculation yield a non-zero magnetization above rather small values of $U$. However, the integrated local moment $M=\sum_{i\in A} m_i$ is far below the quantized value of $M=1\mu_B$. A characteristic snapshot of the magnetization density computed for a simulation cell with 162 atoms within the mean field Hubbard model is shown in Fig.~\ref{MF} (a). The dominant magnetic moments appear in the sublattice opposite to the one of the defect, but small contributions of opposite sign appear in the same sublattice.
The influence of the coupling to infinite graphene is neatly shown in a calculation where we artificially tune the intensity of the interaction between the central simulation cell $A$, and the rest of graphene.
For that matter, we define the following modified full Green's function
\begin{equation}
  \widetilde{G}_{A}^\lambda(E) =
  \frac{1}{E+i\eta-\widetilde{H}_{A}-\lambda_\Sigma\,\Sigma_{AB}(E)}
\end{equation}
where $\lambda_\Sigma\in[0,1]$ is a control parameter that smoothly interpolates
between limit where the region $A$ is decoupled ($\lambda_\Sigma = 0$) from the rest of the universe, quantum dot regime, and the infinite-crystal regime
($\lambda_\Sigma = 1$).

As it is shown in Fig.~\ref{MF} (b), in the quantum dot regime $\lambda_\Sigma = 0$ the magnetic moment is quantized $M=1\mu_B$. However, as soon as the unit cell is coupled to the rest of the graphene, $\lambda_\Sigma \ne 0$, the magnetic moment rapidly becomes non-quantized, with an evolution that depends on the size and geometry of the region $A$.
\begin{figure}[h!]
\centering
\includegraphics[width=0.45\textwidth]{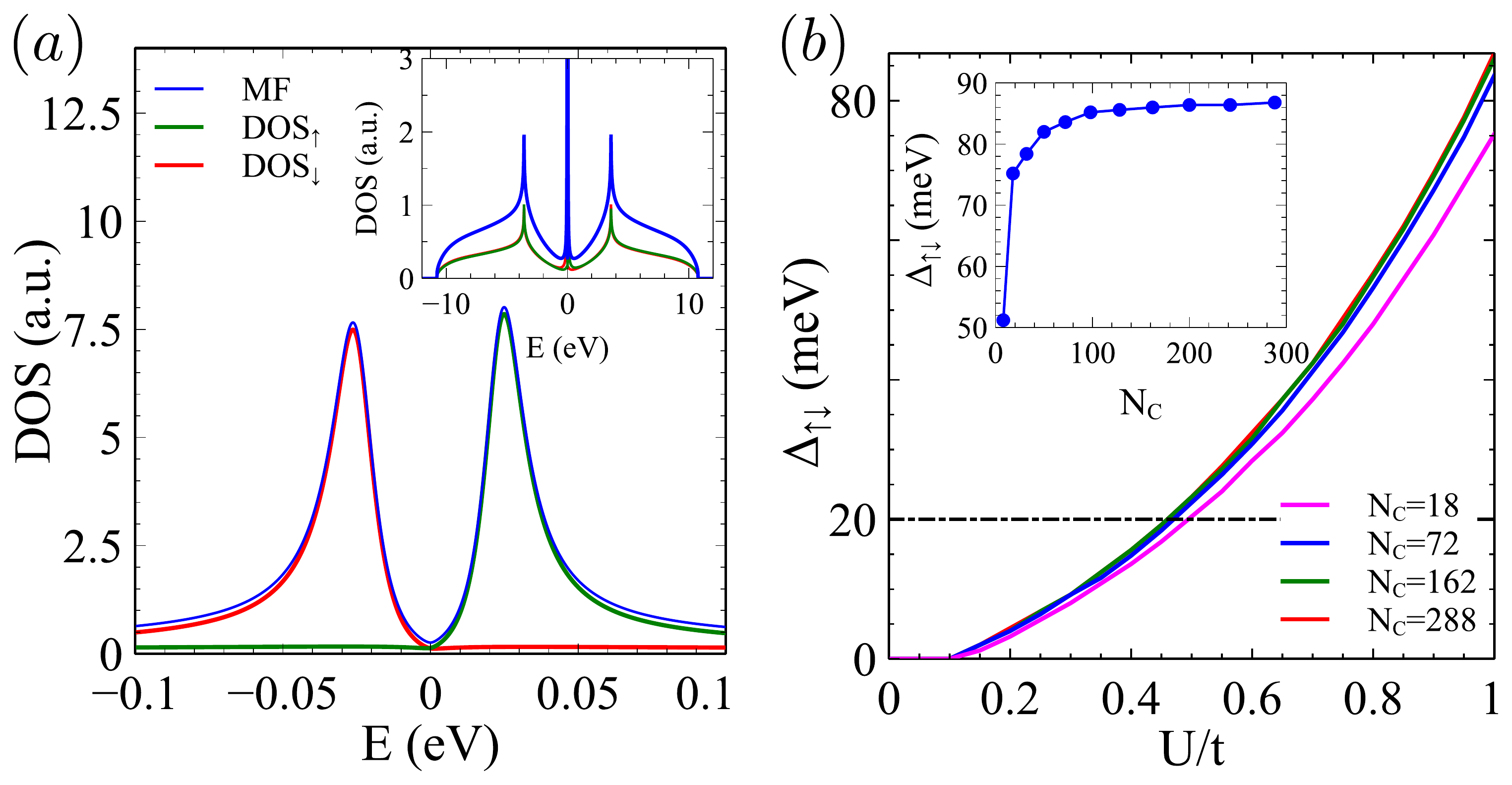}
\vspace{-15pt}
\caption{$(a)$ total and spin-resolved DOS for the Hubbard model. $(b)$ spin splitting, $\Delta$ as a function of the Hubbard interaction $U$ and, in the inset, as a function of the size of the unit cell.}
\label{MFdos}
\end{figure}
As soon as $U$ is larger than a small critical $U\simeq 0.01 t$, magnetic solutions are obtained and we find that the non integer nature of the magnetic moment holds for a wide regime of electronic interactions $U$.
The net magnetic moment is an increasing function of $U$ as well as the size of the central region, $N_C$, Fig.~\ref{MF} (c,d). Even for the largest simulation cells, with up to 288 sites, the total magnetic moment remains clearly below $1\mu_B$. However, a representation of the total moment as a function of $N_C^{-1}$, not shown, makes it hard to predict whether or not the extrapolation to an infinite cell would recover the quantized value. Whereas it might be that the magnetic moment is quantized, our calculations emphasize the rather extended nature of this object.

\subsection{Spin splitting}
Our magnetic self-consistent solutions spontaneously break symmetry and result in spin-split density of states, shown in Fig.~\ref{MFdos}.
Importantly, the interacting DOS does not have any integrable singularity, as it happens for the $U=0$ case at $E=0$. Summing over spin projections, the total density of states still shows electron-hole symmetry.
However, the spin-resolved DOS is split, so for one spin projection the resonance is below the Fermi energy, while for the other one is above, which accounts for the net magnetization.
We define the spin splitting, $\Delta$, as the difference in energy between these two resonances. We find that $\Delta$ has a super-linear dependence on the Hubbard interaction $U$, as shown in Fig.~\ref{MFdos} (b). This reflects the fact that the spin splitting is linear both in $U$ and in the magnetic density $m$, which is also an increasing function of $U$. These results depend weakly on the size of the defective region, $A$, in the embedding calculation (inset of Fig.~\ref{MFdos} (b)).

In order to compare with the experimental observations,\cite{gonzalez2016atomic} that also show two peaks in the DOS with a splitting of $\Delta\sim20meV$ (black dashed line in Fig.~\ref{MFdos} (b)).
However, the interpretation of the experiments requires some cautionary remarks. The image of spin-split peaks would definitely make sense in a magnetic system that breaks symmetry, such as a ferromagnetic system, large enough to keep the magnetization frozen along a given direction that defines a spin quantization axis that permits the definition of the spin orientation of the spin-split bands.
The quantum fluctuations of this mean field picture can be safely neglected for a large enough magnetic moment, but this is definitely not the case of a system that, at most, has $S=1/2$.
Therefore,  and in line with the discussion Gonzalez {\em et al.} \cite{gonzalez2016atomic},
a more correct interpretation of the split peaks is the following. The STM $dI/dV$ is proportional to the spectral function of the graphene.

A proper treatment of the a local moment will probably give two resonances in the spectral function reflecting the addition energy between the empty and singly occupied level, in the case of the lower addition energy, and a second higher energy peak related the difference between the singly and doubly occupied states\cite{Ijas2013}. For a single in-gap level with wave function $\psi_0$, at the mean field level, this difference in addition energies can be related to $U\sum_i|\phi_0|^4$, that relates to the spin splitting of the broken symmetry solution. Thus, the splitting of the peaks observed in the experiment is a Coulomb-Blockade type of phenomena, rather than a symmetry breaking of the spin levels. Of course, this picture entails the existence of an unpaired spin, very much like in the Anderson model \cite{anderson1961localized} and does not preclude the emergence of Kondo effect at very low temperatures.
The proper treatment of the addition energies in this system would involve solving in a many body framework\cite{haase2011magnetic,sofo2012magnetic,mitchell13} the single impurity problem of a resonance in a Dirac bath, which is out of the scope of the present work.\\

\subsection{Localization}
Our calculations show that magnetic moment associated to a hydrogen ad-atom is dellocalized in more than 250 carbon atoms.
In this sense, our calculations highlight the anomalous nature of the resonance due to a single $sp^3$ impurity in contrast to the phenomenology in gapped systems. This behavior arise from the special condition of the DOS, (null only in exactly one point) and the absence of an energy scale able to confine the $E=0$ resonance.

It is worth noting that in real graphene some effects not captured by the first neighbor Hubbard model that can play a relevant role. First, the existence of second neighbor carbon hopping breaks electron-hole symmetry and shifts  the vacancy state away from $E=0$. Second, single hydrogenation introduces an effective on-site energy in the carbon atom that is actually finite, although rather large, which also leads to a displacement of the resonance away from zero energy.
Third, non-local electronic interaction in graphene may have a sizable effect on the magnetic moment.
And finally, spin-orbit coupling would open a gap of around $0.03meV$.
\cite{PhysRevB.80.235431}
Whether any of the previous perturbations would be capable of moving the system to the conventional quantum dot regime would require a careful study with a first principles Hamiltonian, which is out of the scope of the present work.

\section{Conclusions}
\label{sec:Concl}
We have addressed the problem of the local moment formation induced by an individual $sp^3$ functionalization in graphene, with chemisorbed atomic hydrogen as the  main motivation.  We model this within the single orbital Hubbard model, so that the functionalization is modeled as a vacancy in the  honeycomb lattice.  We have shown that the magnetic moment in this system  departs from  the conventionally accepted $m=1\mu_B$ picture. This relates to the fact that the lack of a gap in graphene prevents the existence of a standard in-gap state that can host an unpaired electron. Our calculations show that the local moment induced by $sp^3$ functionalization in otherwise gapless and pristine graphene
 give rise to specific signatures in the magnetic response of the system, such as a non-Curie temperature dependence and a non-linear (and non-monotonic for
some doping values) magnetic susceptibility.
We have also shown by means of mean-field calculations that the resulting magnetic moment is non-quantized in the whole regime explored. Our results should pave the way for future work treating many-body spin fluctuations beyond mean field theory\cite{haase2011magnetic,sofo2012magnetic,mitchell13}

\section{Acknowledgments}

The authors acknowledge financial support by Marie-
Curie-ITN 607904-SPINOGRAPH. JFR acknowledges financial  supported  by  MEC-Spain  (FIS2013-47328-C2-2-P  and  MAT2016-78625-C2)  and  Generalitat  Valenciana  (ACOMP/2010/070),  Prometeo,  by  ERDF  funds
through the Portuguese Operational Program for Competitiveness  and  Internationalization  COMPETE  2020,
and National Funds through FCT- The Portuguese Foundation  for  Science  and  Technology,  under  the  project
PTDC/FIS-NAN/4662/2014  (016656).   This  work  has
been financially supported in part by FEDER funds. JLL
and  N.G.M.  thank  the  hospitality  of  the  Departamento  de
Fisica Aplicada at the Universidad de Alicante.
J.L.L. thanks P. San-Jose for showing
the kernel polynomial method.
We acknowledge Ivan Brihuega, Manuel Melle-Franco, Juanjo Palacios and H. Gonzalez for fruitful discussions.

\bibliographystyle{apsrev4-1}
\bibliography{biblio}
\end{document}